\documentstyle[psfig,floats,aps,aipbook]{revtex}
\begin{document}
\title{A 23 GHz Survey of GRB Error Boxes}
\author{J.~N. Hewitt$^1$, C.~A. Katz$^1$, S.~D. Barthelmy$^2$, W.~H. Baumgartner$^1$,
T.~L. Cline$^2$, B.~E. Corey$^3$, G.~J. Fishman$^4$, N.~Gehrels$^2$, K.~C. Hurley$^5$,
C. Kouveliotou$^4$, C.~A. Meegan$^4$, C.~B. Moore$^1$, R.~E. Rutledge$^6$, and C.~S. Trotter$^1$}
\address{$^1$ Department of Physics and Research Laboratory of Electronics,
Massachusetts Institute of Technology \\
$^2$ NASA Goddard Space Flight Center \\
$^3$ Haystack Observatory \\
$^4$ NASA Marshall Space Flight Center \\
$^5$ Space Sciences Laboratory, University of California, Berkeley \\
$^6$ Department of Physics and Center for Space Research, Massachusetts
Institute of Technology}
\maketitle
\begin{abstract}
The Haystack 37-meter telescope was used in a pilot project in May 1995 to
observe GRB error boxes at 23~GHz.  Seven BATSE error boxes and two
IPN arcs were scanned by driving the beam of
the telescope rapidly across their area.  For the BATSE error boxes,
the radio observations took place two to eighteen days after the BATSE
detection, and several boxes were observed more than once.
Total power data were recorded continuously as the telescope was driven at
a rate of 0.2~degrees/second, yielding Nyquist sampling of the beam with
an integration time of 50~milliseconds, corresponding to a theoretical
rms sensitivity of 0.5~Jy.
Under conditions of good weather, this sensitivity was achieved.
In a preliminary analysis of the data we detect only two sources,
3C273 and 0552+398, both catalogued sources that are known to be
variable at  23~GHz.
Neither had a flux density that was unusally high or low at the time of our
observations.
\end{abstract}

\section*{Introduction}

The detection of radio counterparts to gamma-ray bursts would provide
important information on the source emission mechanism, would
be likely to produce more accurate position determinations,
and would give a measure of the source distance through a measurement of
the dispersion delay (Palmer 1993).
There is, therefore, considerable interest in detecting radio counterparts
to gamma ray bursts, and several searches are under way.
The Haystack telescope\footnote{Radio astronomy at Haystack Observatory
of the Northeast Radio Observatory Consortium (NEROC) is supported by
the National Science Foundation.}
can make a unique contribution because of its high frequency capabilities
and its ability to slew rapidly.

The Haystack telescope is optimized for operation above 1~GHz; a
recently completed upgrade  (Barvainis {\it et al.} 1993)
has extended operations to frequencies as high as 116~GHz.
The telescope was originally designed to track objects in low
earth orbit, and high slew speeds can be tolerated.
Any source position above the horizon can be reached from zenith
within 30 seconds, and the complementary goals of high resolution
and wide field of view may be achieved through rapid scanning of an
area of the sky.

\section*{Observations}

Rapid continuum mapping of large fields with the Haystack telescope
had not been previously attempted, and a two-week period in May 1995
was allocated to us to develop the observing techniques.
At the observing frequency of 23~GHz, the beam is 1.4 arc~minutes and the
gain is 0.14~K/Jy.
The observing frequency was chosen as a compromise between the desire
to observe at as high a frequency as possible and a desire to produce a map
within a reasonable period of time.
The bandwidth was 160~MHz and the nominal system temperature was 140~K;
the actual system temperature depended on the elevation and weather
conditions (see below).
Error boxes were scanned at $0.2^\circ$/second in right ascension,
stepping in increments of $0.00625^\circ$ in declination.
Continuum data were sampled at 20~Hz, yielding Nyquist sampling in the
right ascension direction and twice Nyquist sampling in the declination
direction.
Oversampling in declination was necessary because with the rapid scanning
the telescope failed to settle to its commanded declination before
beginning to scan in right ascension.

Seven BATSE error boxes and two IPN arcs were observed; the observations
are summarized in Table~1.
Each error box was scanned in several ``slices;'' and in most cases
each slice was observed more than once.
Observing conditions varied considerably with source elevation and as
the weather changed.
For presentation in Table~1 we selected just one observation of each
slice; the observation selected was that with the best sensitivity.

%
\begin{table}
\caption{Journal of Observations}
\label{tableone}
\begin{tabular}{l l c c c}
& {\bf Date and} & {\bf Date of} & {\bf Theoretical rms} &  \\
& {\bf Time of}  & {\bf Radio} &  {\bf Sensitivity} & {\bf Size of} \\
& {\bf Burst (UT)} & {\bf Observations (UT)} & {\bf (Jy)} & {\bf Map} \\
\tableline
IPN3509\_1 & 95/04/16 & 95/05/24 & 0.52 & $4.4^\circ \times 0.6^\circ$ \\
IPN3509\_2 & ~13:26:59.76& 95/05/22 & 0.52 & $4.3^\circ \times 0.8^\circ$ \\
IPN3509\_3 &          & 95/05/22 & 0.52 & $3.7^\circ \times 1.0^\circ$ \\
IPN3509\_4 &          & 95/05/22 & 0.53 & $2.6^\circ \times 1.0^\circ$ \\
\noalign{\vskip 0.2 true in}
IPN3512\_1 & 95/04/18 & 95/05/21 & 0.58 & $1.0^\circ \times 1.5^\circ$ \\
IPN3512\_2 & ~23:16:35.63 & 95/05/21 & 0.58\\
IPN3512\_3 &          & 95/05/21 & 0.58\\
IPN3512\_4 &          & 95/05/21 & 0.57\\
IPN3512\_5 &          & 95/05/21 & 0.59\\
IPN3512\_6 &          & 95/05/22 & 0.51\\
IPN3512\_7 &          & 95/05/22 & 0.53\\
IPN3512\_8 &          & 95/05/22 & 0.56\\
IPN3512\_9 &          & Not observed \\
IPN3512\_10 &         & 95/05/22 & 0.58 \\
IPN3512\_11 &          & 95/05/21 & 0.58 \\
IPN3512\_10 &          & 95/05/21 & 0.58 \\
\noalign{\vskip 0.2 true in}
BATS3552\_1 & 95/05/06 & 95/05/27 & 0.51 & $5.7^\circ \times 1.3^\circ$ \\
BATS3552\_2 & ~23:25:12.75& 95/05/27 & 0.50 \\
BATS3552\_3 &          & 95/05/28 & 0.52 \\
BATS3552\_4 &          & 95/05/28 & 0.51 \\
BATS3552\_5 &          & 95/05/28 & 0.50 \\
\noalign{\vskip 0.2 true in}
BATS3567\_1 & 95/05/09 & 95/05/24 & 0.79 & $6.4^\circ \times 1.1^\circ$\\
BATS3567\_2 & ~23:16:05.93& 95/05/24 & 0.62 \\
BATS3567\_3 &          & 95/05/24 & 0.58 \\
BATS3567\_4 &          & 95/05/23 & 0.53 \\
BATS3567\_5 &          & 95/05/24 & 0.57 \\
BATS3567\_6 &          & 95/05/23 & 0.49 \\
\noalign{\vskip 0.2 true in}
BATS3588\_1 & 95/05/21 & 95/05/23 & 0.68 & $4.7^\circ \times 0.9^\circ$ \\
BATS3588\_2 & ~06:59:17.29& 95/05/23 & 0.65 \\
BATS3588\_3 &          & 95/05/23 & 0.68 \\
BATS3588\_4 &          & 95/05/23 & 0.65 \\
BATS3588\_5 &          & 95/05/23 & 0.67 \\
BATS3588\_6 &          & 95/05/23 & 0.65 \\
\noalign{\vskip 0.2 true in}
BATS3593\_1 & 95/05/22 & 95/05/26 & 0.52 & $3.4^\circ \times 0.85^\circ$ \\
BATS3593\_2 & ~23:41:23.18  & 95/05/26 & 0.50 \\
BATS3593\_3 &          & 95/05/31 & 0.53 \\
BATS3593\_4 &          & 95/05/30 & 0.56 \\

\end{tabular}
\end{table}

\setcounter{table}{0}
\begin{table}
\caption{continued}
\begin{tabular}{llccc}
\noalign{\vskip 0.2 true in}
BATS3594\_1 & 95/05/23 & 95/05/27 & 0.53 & $6.6^\circ \times 1.1^\circ$ \\
BATS3594\_2 & ~05:39:29.07  & 95/05/27 & 0.54 \\
BATS3594\_3 &          & 95/05/27 & 0.50 \\
BATS3594\_4 &          & 95/05/28 & 0.65 \\
BATS3594\_5 &          & 95/05/27 & 0.50 \\
BATS3594\_6 &          & 95/05/28 & 0.52 \\
\noalign{\vskip 0.2 true in}
BATS3598\_1 & 95/05/24 & 95/05/26 & 0.55 & $4.2^\circ \times 1.1^\circ$ \\
BATS3598\_2 &  ~04:08:19.12 & 95/05/26 & 0.51 &  \\
BATS3598\_3 &          & 95/05/27 & 0.52 &  \\
BATS3598\_4 &          & 95/05/27 & 0.55 &  \\
\tablenotetext{The theoretical rms sensitivity is given by $T_{\rm sys} / (\sqrt{\Delta \nu \tau} G)$
where $T_{\rm sys}$ is the measured system temperature at the time
of observation, $\Delta \nu = 160$~MHz, $\tau = 50$~msec, and
$G = 0.14$~K/Jy.}
\end{tabular}
\end{table}

The data were displayed and inspected in real-time during the observations
and we believe we would have noted any source with flux density exceeding
ten times the rms sensitivity listed in Table~1.
This detection limit must be corrected for the weather-dependent opacity
at 23~GHz; for the data presented here the correction is typically about
15\%.

As an example of the quality of the data produced in the survey we present
in Figure~1 our map of part of the error box of BATSE~\#3598.
This map was produced by fitting a linear baseline to five-second segments of
data and subtracting the fit from the central three seconds of data.
The measured pixel-to-pixel rms was 0.49~Jy, consistent with the expected
value of 0.51~Jy (see Table~1).
Regions in which the fit was poor are evident in the map as
horizontal ``streaks'' over a portion of a single row; occasional
unsampled pixels can also be seen.
Sources (here, 3C273) appear as bright regions with the extent of the beam.
Further data analysis is in progress.

%
\begin{figure}
\psfig{file=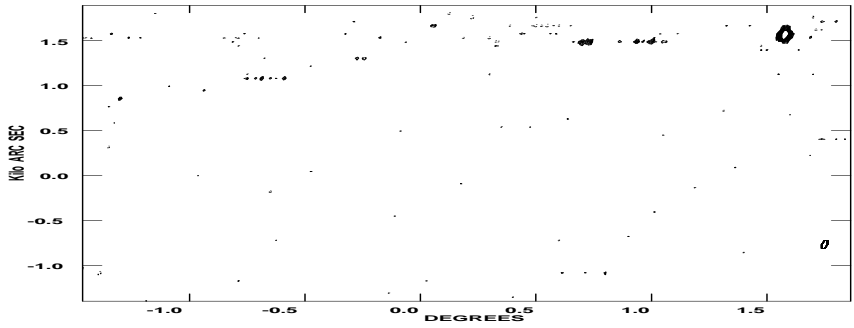,height=2in,width=4.5in}
\caption{Contour plot of map of part of the error box of BATSE\#3598.
The contours are -3, 3, 4, 5, 6, 7, 8, 9, 10, 11, 12, 13, 14, 15,
16, 17, 18, 19, and 20 times the root-mean-square value of the map
of 0.49~Jy.  The source visible in the upper right corner is 3C273.}
\label{3c273slice}
\end{figure}

\section*{Results}

In the preliminary analysis of the data during the observations, only 3C273
was detected; 0552+398 was noted when we examined our maps for previously
known, catalogued sources.
No other source was detected in the maps summarized in Table~1.
3C273 is in the error box of BATSE\#3598 (see Figure~1);
0552+398 is in the error box of BATSE\#3594.
The data analysis is still in progress; careful examination of all radio
maps and averaging of the data for boxes scanned more than once may 
yield further detections.

It is possible that radio-loud AGN are the sources of the gamma ray bursts.
However, both 3C273 and 0552+398 are already known to be variable and neither
had a flux density that was unusually high or low during the time of our 
observations.  Therefore, there is no compelling reason to associate
either with a GRB.

\section*{Future Prospects}

Now that the continuum mapping technique with the Haystack telescope
has been demonstrated, further observations of GRB error boxes have
been proposed.
During this past summer, the telescope servo system was replaced.
With the improved pointing control more rapid scanning of boxes should
be possible, making feasible observations at higher frequencies.
A new receiver will provide lower receiver noise at 23~GHz and allow us
to observe farther from the 22~GHz water line, further increasing the
sensitivity.


\end{document}